\documentclass[a4paper,fleqn,usenatbib]{mnras}

\usepackage{multirow}
\usepackage{graphicx}
\usepackage{amsmath}
\usepackage{epsfig}
\usepackage{color}
\usepackage{amssymb}
\usepackage{float}

\usepackage{mathptmx}
\usepackage{txfonts}

\usepackage[T1]{fontenc}
\usepackage{ae,aecompl}

\usepackage{bmpsize}

\title[Mass and Metallicity Requirement in Stellar Models]
{Mass and Metallicity Requirement in Stellar Models for Galactic Chemical Evolution Applications}
\author[C\^ot\'e et al.]
{Benoit C\^ot\'e,$^{1,2,3,10,11}$
Christopher West,$^{4,10}$
Alexander Heger,$^{5,6,10,11}$\newauthor
Christian Ritter,$^{1,10,11}$
Brian W. O'Shea,$^{2,3,7,10}$
Falk Herwig,$^{1,10,11}$\newauthor
Claudia Travaglio$^{8,9,11}$ and
Sara Bisterzo$^{8,10,11}$
\\
$^{1}$Department of Physics and Astronomy, University of Victoria, Victoria, BC, V8W 2Y2, Canada\\
$^{2}$Department of Physics and Astronomy, Michigan State University, East Lansing, MI, 48824, USA\\
$^{3}$National Superconducting Cyclotron Laboratory, Michigan State University, East Lansing, MI, 48824, USA\\
$^{4}$Center for Academic Excellence, Metropolitan State University, St Paul, MN, 55106, USA\\
$^{5}$Monash Centre for Astrophysics, Monash University, Melbourne, Victoria, 3800, Australia\\
$^{6}$Center for Nuclear Astrophysics, Department of Physics and Astronomy, \\ \,\,\,Shanghai Jiao-Tong University, Shanghai, 200240, P. R. China\\
$^{7}$Department of Computational Mathematics, Science and Engineering, Michigan State University, East Lansing, MI, 48824, USA\\
$^{8}$INAF, Astrophysical Observatory Turin, Strada Osservatorio 20, I-10025, Pino Torinese (Turin), Italy\\
$^{9}$B2FH Association, Turin, Italy\\
$^{10}$Joint Institute for Nuclear Astrophysics - Center for the Evolution of the Elements, USA\\
$^{11}$NuGrid Collaboration, \href{http://nugridstars.org}{http://nugridstars.org}
}
\begin{document}

\date{Accepted XXX. Received XXX; in original form XXX}

\pubyear{2016}

\pagerange{\pageref{firstpage}--\pageref{lastpage}} \pubyear{XXX}

\maketitle

\label{firstpage}

\begin{abstract}
We used a one-zone chemical evolution model to address the question
of how many masses and metallicities are required in grids of 
massive stellar models in order to ensure reliable galactic chemical evolution predictions.
We used a set of yields that
includes seven masses between 13 and $30\,$M$_\odot$, 15
metallicities between 0 and 0.03 in mass fraction, and two different
remnant mass prescriptions.  We ran several simulations where we
sampled subsets of stellar models to explore the impact of different
grid resolutions. Stellar yields from low- and intermediate-mass
stars and from Type~Ia supernovae have been included in our simulations,
but with a fixed grid resolution.  We compared our results with the stellar abundances
observed in the Milky Way for O, Na, Mg, Si, Ca, Ti, and Mn.
Our results suggest that the range of metallicity considered is more
important than the number of metallicities within that range, which
only affects our numerical predictions by about 0.1~dex.  We found
that our predictions at [Fe/H]~$\lesssim-2$ are very sensitive to
the metallicity range and the mass sampling used for the lowest
metallicity included in the set of yields.  Variations between results
can be as high as 0.8 dex.
At higher [Fe/H], we found that the required number of masses depends on the element of interest
and on the remnant mass prescription.  With a monotonic remnant mass
prescription where every model explodes as a core-collapse supernova,
the mass resolution induces variations of 0.2~dex on average.  But
with a remnant mass prescription that includes islands of
non-explodability, the mass resolution can cause variations of about
0.2 to 0.7~dex depending on the choice of the lower limit of the
metallicity range.  With such a remnant mass prescription, explosive
or non-explosive models can be missed if not enough masses are
selected, resulting in over- or under-estimations of the mass ejected
by massive stars.
\end{abstract}


\section{Introduction}
\begin{figure}
\begin{center}
\includegraphics[width=3.6in]{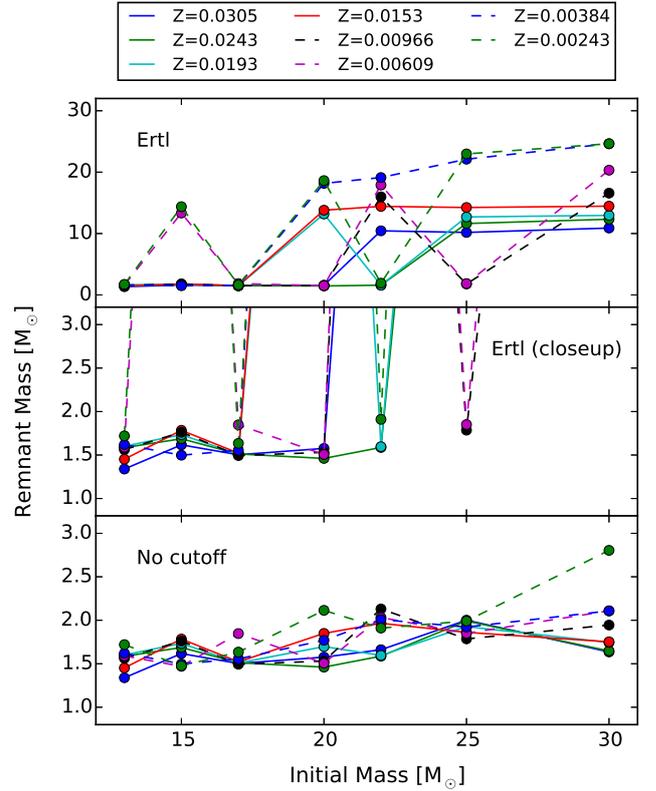}
\caption{Remnant mass as a function of stellar initial mass for the
  eight highest metallicities available in the yields described in
  Section~\ref{sect_yields}, using the \protect\cite{e15} (upper and
  middle panels) and the no-cutoff (lower panel) prescriptions.
  Remnant masses larger than~$\sim3\,$M$_\odot$ implies the formation
  of a black hole.  $[Z]=\mathrm{log}_{10}(Z/Z_\odot)$ where $Z_\odot=0.0153$.}
\label{fig_remnant_high}
\end{center}
\end{figure}

\begin{figure}
\begin{center}
\includegraphics[width=3.5in]{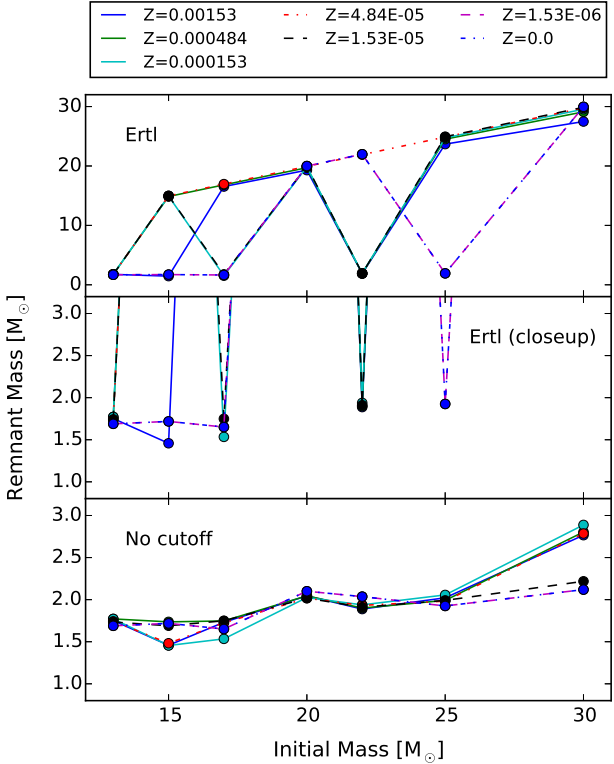}
\caption{Same as Figure~\ref{fig_remnant_high} but for the 7 lowest metallicities available in the yields described in Section~\ref{sect_yields}.}
\label{fig_remnant_low}
\end{center}
\end{figure}

Stellar yields are fundamental ingredients in chemical evolution
models and simulations.  To reproduce the chemical enrichment of
galaxies over their entire lifetime, those yields need to include
low-mass, intermediate-mass, and massive stars along with a wide range
of metallicities, ideally from zero metallicity up to solar
composition.  Several grids of stellar models are available in the
literature with different numbers of masses and metallicities.  In
general, for massive stars, those grids either offer a limited number
of masses within a certain range of metallicities
(e.g., \citealt{ww95,pcb98,cl04,k06,p13})
or a large number of masses for one
specific metallicity (e.g., \citealt{lc06,wh07,hw10,e12}).\\
\indent In this paper we address the question of how many masses and
metallicities for massive stars are required in a grid of stellar models to ensure
convergence in galactic chemical evolution studies.  To do so, we
conduct an experiment with a set of yields which
has seven masses from 13 to 30\,M$_\odot$ and 15
metallicities (see Section~\ref{sect_yields}), where we sample subsets of stellar
models to create new sets of yields with lower mass and metallicity
resolutions.  We then fold those yields into simple stellar
populations (SSPs) and include them in a one-zone chemical evolution
model to quantify the impact of the stellar grid resolution on our
predictions.  For now, we do not consider the impact of 
massive binary systems which could  significantly alter the evolution
and the ejecta of massive stars (e.g., \citealt{ddv04,s12,dm13}).  Although our sensitivity study
only focuses on massive stars, we included the contribution of low-and
intermediate-mass stars and Type~Ia supernovae (SNe~Ia) in our calculations (see Section~\ref{sect_yields}).\\
\indent It is generally believed that the most massive stars are more likely
to form a black hole and lock away most of the heavy elements
synthesized during their evolution (e.g., \citealt{whw02,h03,zwh08}).
This seems to be also supported by the
lack of observed more massive progenitors for common supernovae
(e.g., \citealt{s09,w14}).  This sustains the idea that there must be a
transition mass above which massive stars stop to contribute to the
chemical evolution of galaxies (but see discussion in Section~\ref{sect_yields}).
Recent studies with high mass resolutions,
however, suggest that such a transition may not exist and that black
hole formation is in fact sparsely distributed across the stellar
initial mass spectra, forming islands of non-explodability 
(e.g., \citealt{u12,e15,s15}).\\
\indent The yields used in this work for massive stars have been
calculated with four different remnant mass and black hole formation
prescriptions, which enables us to study the impact of such
prescriptions on our grid resolution study.  As two extreme cases, we
consider the prescription of \cite{e15} that generates islands of
non-explodability, and the \textit{no-cutoff} prescription, which is a
monotonic remnant mass distribution where all models explode with
minimum fallback (see Figures~\ref{fig_remnant_high} and
\ref{fig_remnant_low}).  The remnant mass is the baryonic final mass of
a star (i.e., not its gravitational mass) and refers to its initial
mass minus the total mass ejected during its lifetime, which includes
the explosive ejecta, if any, and stellar winds.  In our specific
case, any remnant mass larger than~$\sim3\,$M$_\odot$ implies a black
hole formation instead of an explosion.  We also assumed that, if a
star does not explode, the entire star disappears as a back hole
and no supernova yield is produced.  Whereas this may be a simplification in
some cases, e.g., the formation of long-duration gamma-ray bursts or
some types of hypernovae, it should be reasonable to assume that in
this case the bulk, and in particular the inner parts of the stars,
are not being ejected.

Throughout this paper, we compare our numerical predictions with
observations to provide a visual reference to evaluate the
importance of the stellar grid resolution in our simulations.  We chose the
Milky Way because of the large amount of stellar abundances data and
because of the wide metallicity range covered by those data.  Although
one-zone models do not capture the complexity of the formation of
massive systems such as the Milky Way, we believe they are sufficient,
at least as a first order approximation, to address the specific
question of what is the impact of the stellar grid resolution and the
remnant mass prescription in the context of galactic chemical
evolution.  Our results may differ from the ones generated 
by two-zone and three-zone models (e.g., \citealt{f92,pfm95}),
since all of our stellar ejecta is returned and recycled in
a unique gas reservoir instead of being distributed within the
different galactic structures, such as the halo and the thick and thin discs.
It is not our goal to produce the most realistic model of the
Milky Way.  More sophisticated simulations for our Galaxy can be found
in the literature (e.g., \citealt{cmr01,t04,kn11,mmr13,m13,mcgg15,shen15,vv15,w15}).

This paper is organized as follow.  In Section~\ref{sect_gal_model}, we describe our
chemical evolution code and input physics.  We describe our stellar abundances data
selection in Section~\ref{sect_sad}.  The impact of the mass
and metallicity resolutions is presented in Section~\ref{sec_y_mrm} for stellar yields with
monotonic remnant masses, and in Section~\ref{sect_y_ine} for stellar yields with islands
of non-explodability.  We summarize our results and give our conclusions in Section~\ref{sect_s_c}.

\section{Galaxy Model}
\label{sect_gal_model}
We use OMEGA, our One-zone Model for the Evolution of GAlaxies code
(\citealt{c16}), to follow the chemical evolution of the
Milky Way.  The input parameters of the closed-box version and the
treatment of SSPs using SYGMA, which stands for Stellar Yields for
Galactic Modeling Applications (C. Ritter et al.,~in prep.), are described
in details in \cite{c15}.  For the present study, we consider an
open box model that includes and inflows of primordial gas and galactic outflows.
All of our codes are available online with the NuGrid NuPyCEE
package\footnote{\href{https://github.com/NuGrid/NUPYCEE}{https://github.com/NuGrid/NUPYCEE}}.

\subsection{Open Box Model}
\label{sect_obm}
OMEGA uses the classical equations of single-zone chemical evolution
models (\citealt{p09}).  At a certain time, $t$, the mass of the gas
reservoir, $M_{\mathrm{gas}}$, is calculated by
\begin{equation}
M_{\mathrm{gas}}(t+\Delta t)=M_{\mathrm{gas}}(t) + \Big[\dot{M}_{\mathrm{in}}(t) - \dot{M}_{\mathrm{out}}(t) +\dot{M}_{\mathrm{ej}}(t) - \dot{M}_{\mathrm{\star}}(t)\Big]\Delta t,
\label{eq_main}
\end{equation}

\noindent where $\Delta t$ is the length of the timestep and the four terms in brackets are, from left to right, the inflow rate, the outflow rate, the stellar mass loss rate of all the SSPs, and the star formation rate.  The mass ejected by each SSP is calculated using the initial mass function of \cite{c03} and the stellar yields described in Section~\ref{sect_yields}.  We use a star formation history that has a similar shape than the one derived from the two-infall Milky Way model of \cite{cmg97,cmr01}, which we normalized to produce a current stellar mass of $5\times10^{10}\,$M$_\odot$ (\citealt{f06,mcm11,bv13,ln15}) at the end of our simulations.

\subsection{Outflow Rate}
The evolution of the galactic outflow rate is defined by (\citealt{mqt05})
\begin{equation}
\dot{M}_{\mathrm{out}}(t) = \eta(t) \dot{M}_{\mathrm{\star}}(t).
\end{equation}
To calculate the time evolution of $\eta$, the mass-loading factor, we use
the \textit{MA} prescription described in \cite{c16},
\begin{equation}
\label{eq_eta_z}
\eta(z)\propto M_\mathrm{vir}(z)^{-\gamma/3}(1+z)^{-\gamma/2},
\end{equation}
where $M_\mathrm{vir}$ is the total virial mass of the system.   The redshift, $z$, is
converted into time using the cosmological parameters measured in \cite{d09},
assuming the end of our simulations represents $z=0$.  We set $\gamma$ to unity
to consider outflows driven by radiative pressure (see \citealt{mqt05}).
We assume the dark matter halo mass of the Milky Way follows the
equations derived by \cite{fak10} which represents the average dark
matter accretion rates extracted from the Millennium simulations (\citealt{s05,bk09}).
In each series of simulations, we tune the final value of the mass-loading
factor to ensure that the peak of the predicted metallicity distribution function
occurs at [Fe/H]~$\sim0$.

\subsection{Mass of Gas and Inflow Rate}
At every time $t$, we assume the star formation follows the
Kennicutt-Schmidt law (\citealt{s59,k98}) in the adapted form of
(\citealt{baugh06,sd15})
\begin{equation}
\label{eq_SF_law}
\dot{M}_\mathrm\star(t)=f_\star\frac{M_\mathrm{gas}(t)}{\tau_\mathrm\star(t)},
\end{equation}

\noindent where $f_\mathrm\star$ and $\tau_\mathrm\star$ are respectively the star
formation efficiency and the star formation timescale.  Because
$\dot{M}_\star$ is a known quantity in our code, we can reverse
equation (\ref{eq_SF_law}) and derive the evolution of the mass of gas
as a function of time.  The inflow rate, the only unknown in
equation~(\ref{eq_main}), can then be isolated and calculated for each
timestep.  This approach has been used in previous works to calculate
the chemical evolution of local dwarf spheroidal galaxies (\citealt{fggl06,g07,h15}).

\subsection{Star Formation Efficiency and Timescale}
\label{sect_sfet}
We assume that the star formation timescale is proportional to the
dynamical timescale, $\tau_\mathrm{dyn}$, of the virialized system
hosting the galaxy (e.g. \citealt{kcdw99,clbf00,swtk01}), and is
defined by $\tau_\star = f_\mathrm{dyn} \tau_\mathrm{dyn}\approx~f_\mathrm{dyn}R_\mathrm{vir}/V_\mathrm{vir}$,
where $f_\mathrm{dyn}$ is the proportional constant and
$R_\mathrm{vir}$ and $V_\mathrm{vir}$ are respectively the
virial radius and the circular velocity of the system. With the
relation for $R_\mathrm{vir}$ defined in \cite{wf91},
\begin{equation}
R_\mathrm{vir}=0.1H_0^{-1}(1+z)^{-3/2}V_\mathrm{vir},
\end{equation}
where $H_0$ is the current Hubble constant, the dynamical timescale is then
given by
\begin{equation}
\label{eq_tau_z}
\tau_\mathrm{dyn}=0.1H_0^{-1}(1+z)^{-3/2}\;.
\end{equation}

With our set of equations, the $f_\mathrm\star/f_\mathrm{dyn}$ ratio is
used to control the initial and final mass of gas in our
simulations.  The initial mass of gas sets the speed and the
concentration of the early enrichment and therefore the metallicity at
which SNe~Ia start to contribute to the chemical evolution, whereas
the final mass of gas sets the final metallicity and the fraction of
gas converted into stars.  We fixed $f_\star/f_\mathrm{dyn}$ to
$0.4$ so that the final mass of gas in our simulated galaxy is
$\sim10^{10}$\,M$_\odot$, consistent with the current state
of the Milky Way (see Table~1 in \citealt{kpa15}).  This represents a star formation
efficiency of 0.04 when $f_\mathrm{dyn}=0.1$.  This choice, however,
implies a relatively low initial mass of gas and generates a fast
early enrichment that pushes the appearance of SNe~Ia up to [Fe/H]
of $\sim-0.5$, which is too high compared to the canonical value of $-1.0$
constrained by observations (see \citealt{mg86,cmr01}).

To solve this issue, we introduced a free parameter, $\mu$, in the exponent
of the redshift dependency term of equation (\ref{eq_tau_z}) so that the
star formation timescale is now described by
\begin{equation}
\label{eq_t_star_mu}
\tau_\star=0.1f_\mathrm{dyn}H_0^{-1}(1+z)^{-3\mu/2}\;.
\end{equation}
This assumes that the gas fraction in our galaxy model does not
necessarily scale linearly with the dynamical timescale of the virialized system.  
It allows us to control the growth of the gas content and to tune the initial mass of
gas independently of the final
mass of gas to make sure that SNe~Ia occur at [Fe/H]~$\sim-1$.
The value of $\mu$ depends on the choice of stellar yields and the
amount of Fe ejected by massive stars.  We recall that our one-zone
model is mostly designed to mimic the evolution of known galaxies
rather than to study how that evolution is driven. Although the $\mu$
parameter has been introduced for fine-tuning, we believe it is
necessary in order to recover the global
properties of the Milky Way with our simple model.  It allows us
to apply our chemical evolution calculations on top of a reasonable gas evolution pattern.

\begin{figure}
\begin{center}
\includegraphics[width=3.7in]{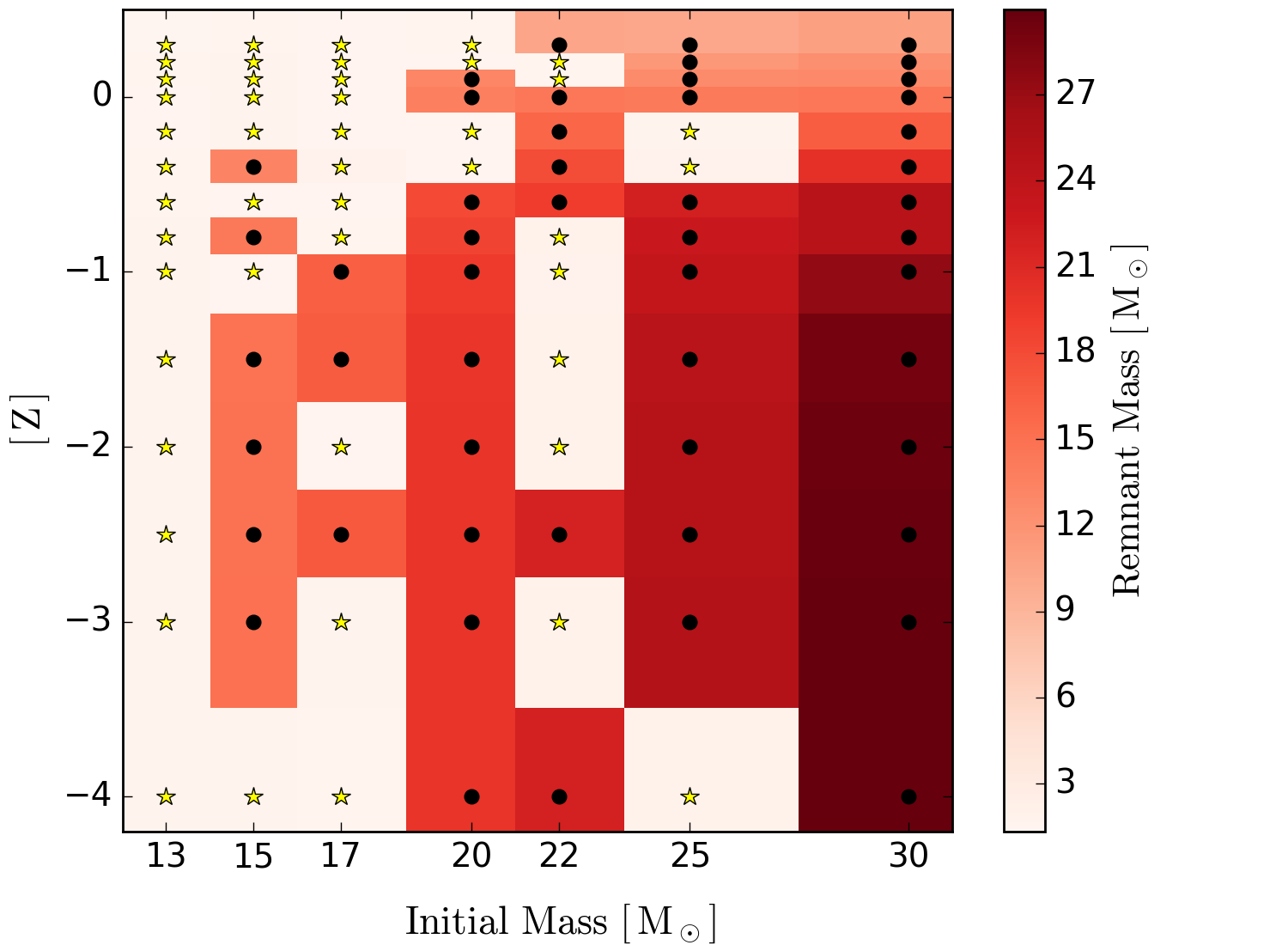}
\caption{Remnant mass as a function of stellar initial mass and metallicity for the \protect\cite{e15} prescription.
Yellow stars and black dots represent explosive and non-explosive models, respectively.}
\label{fig_2D_remnant}
\end{center}
\end{figure}

\subsection{Stellar Yields}
\label{sect_yields}
Using the KEPLER stellar evolution, nucleosynthesis, and supernova
code (\citealt{wzw78,r02}),
we computed a set of non-rotating stellar models and their
nucleosynthesis yields with seven different initial masses (see
Table~\ref{tab_sample}) and for 15 different metallicities of [$Z$] $=$
0.3, 0.2, 0.1, 0.0, $-0.2$, $-0.4$, $-0.6$, $-0.8$, $-1.0$, $-1.5$, $-2.0$, $-2.5$,
$-3.0$, $-4.0$, and $Z=0$, where $[Z]=\mathrm{log}_{10}(Z/Z_\odot)$
and $Z_\odot=0.0153$.  In this work, however, we do not include [$Z$]~$=$~0.3 
since our numerical predictions do not reach such high metallicity. Initial abundances were taken from the
galactic chemical history model of \cite{wh13}.  
During the hydrostatic burning phases, mass loss is treated using the
\cite{ndj90} rate, taking into account the metallicity-dependence
with a power law of exponent 0.5.  All models were
exploded using a flat explosion energy of 1.2 B (1 B $= 10^{51}\,$erg)
for the final kinetic energy of the ejecta due to the lack of
predictive power of current best supernova explosion models for the
explosion energy in the mass range studied here.  Supernova fallback
was obtained self-consistently using the 1D hydro of KEPLER.  We also
assume a \textit{standard} amount of mixing during the supernova explosion
as in which was adjusted to match supernova light curves (\citealt{r02}).  Detail
of this grid will be published in West \& Heger~(in prep.); the $Z=0$ models are
from \cite{hw10}.

\begin{table}
\caption{Mass and metallicity samples extracted from the original set
  of yields described in Section~\ref{sect_yields}.  The nomenclature
  is used in all the figures presented in this study.} \centering
\begin{tabular}{cc}
\hline
Nomenclature & Sample from the original grid\\
\hline
\noalign{\medskip}
\textbf{Mass [M$_\odot$]} &  \\
\noalign{\medskip}
7 M & All masses (13, 15, 17, 20, 22, 25, 30) \\
 \noalign{\medskip}
4 M A & 13, 15, 20, 25 \\
 \noalign{\medskip}
4 M B & 13, 17, 22, 30 \\
\noalign{\medskip}
\hline
\noalign{\medskip}
\textbf{Metallicity} & \\
\noalign{\medskip}
14 Z & All $[Z]$ except 0.3 (see Figures~\ref{fig_remnant_high} and \ref{fig_remnant_low}) \\
 \noalign{\medskip}
6 Z & $Z=0.0$, $[Z]=-2$, $-1$, $-0.4$, $-0.2$, $0.1$ \\
 \noalign{\medskip}
5 Z & $[Z]=-2$, $-1$, $-0.4$, $-0.2$, $0.1$ \\
\noalign{\medskip}
\hline
\label{tab_sample}
\end{tabular}
\end{table}

\begin{table*}
\caption{List of free parameters and final properties in our Milky Way models.
The second column shows the current values observed in the Milky Way disc (see references in \protect\citealt{kpa15}).
The third and fourth columns represent our results using our two different remnant
mass prescriptions.} \centering
\begin{tabular}{cccc}
\hline
 & Milky Way & \multicolumn{2}{c}{OMEGA}\\
 & \cite{kpa15} & No cutoff & \cite{e15}\\
\hline
\noalign{\medskip}
\textbf{Parameters} & & & \\
\noalign{\medskip}
$\eta$, mass-loading factor (see equation \ref{eq_eta_z})& --- & 0.25 & 0.0 \\
 \noalign{\medskip}
$f_\mathrm\star/f_\mathrm{dyn}$, star formation efficiency (see Section \ref{sect_sfet})& --- & 0.4 & 0.4 \\
 \noalign{\medskip}
$\mu$, growth of gas content (see equation \ref{eq_t_star_mu})& --- & 0.3 & 0.7 \\
\noalign{\medskip}
\hline
\noalign{\medskip}
\textbf{Final properties} & \\
\noalign{\medskip}
Stellar mass$^a$ [$10^{10}$ M\,$_\odot$] & 3.0 - 4.0 & 5.0 & 5.0 \\
\noalign{\medskip}
Gas mass [$10^9$ M\,$_\odot$] & $8.1\pm4.5$ & 9.3 & 9.3 \\
\noalign{\medskip}
Star formation rate [M\,$_\odot$ yr$^{-1}$] & 0.65 - 3 & 2.55 & 2.55 \\
\noalign{\medskip}
Infall rate [M\,$_\odot$ yr$^{-1}$] & 0.6 - 1.6 & 1.4 & 1.2 \\
\noalign{\medskip}
Core-collapse SN rate [per 100 yr] & $2\pm1$ & 2.6 & 2.6 \\
\noalign{\medskip}
Type Ia SN rate [per 100 yr] & $0.4\pm0.2$ & 0.4 & 0.4 \\
\noalign{\medskip}
\hline
\multicolumn{4}{l}{$^a$ See Section \ref{sect_obm} for additional references for the total stellar mass of the Milky Way.}
\label{tab_final_prop}
\end{tabular}
\end{table*}

We then employed different criteria to determine whether a successful
explosion would actually occur, based on different criteria in the
literature for the \textit{explodiability} given the pre-supernova
structure of the star at onset of core collapse.  The first simple
case was to assume all stars explode.  In this case, the entire
non-fallback mass of all stars including winds contribute to the
yields.  Next we explored different prescriptions for explodability
based on formula readily available in the literature.  We used the
compactness parameter of \cite{oo11},
\begin{equation}
\xi_M=\left.\frac{M/\mathrm{M}_\odot}{R(M_\mathrm{bary}=M)/1000\,\mathrm{km}}\right|_{t=\mathrm{bounce}}\;,
\end{equation}
with $M=2.5\,\mathrm{M}_\odot$ and cut-off values of $0.25$ as
suggested by \cite{oo11} and $0.45$ as suggested by \cite{s14}, and
the prescription by \cite{e15} with the normalization to Model
s19.8.  When the criteria for black hole formation was fulfilled,
we assumed the entire star would collapse to a black hole instead of
producing supernova nucleosynthesis.  Only contribution from
mass loss due to winds prior to collapse would be present in this
case.  In cases were no black hole is formed, the full yields as
described at the beginning of this paragraph would be used.\\
\indent As we found that the prescription by \cite{e15} is the most extreme in
the sense that it makes the most black holes (see Figure~\ref{fig_2D_remnant}), we only use this here for
comparison.  In the following, we skip the models using the more dated
prescription of \cite{oo11} for the sake of clarity of the discussion.
This prescription produced only some intermediate results compared to
the two extreme cases considered in the present study.

We combined these massive star yields with the low- and
intermediate-mass models calculated by NuGrid (\citealt{p13}; C. Ritter
et al. in prep.), which are available
online\footnote{\href{http://nugridstars.org/data-and-software/yields/set-1}{http://nugridstars.org/data-and-software/yields/set-1}}.
Although we do not focus on elements significantly produced by those
lower-mass models (e.g., carbon), we decided to include them to
account for their contribution in the amount of hydrogen returned in
the gas reservoir.  The ejecta coming from SNe~Ia is calculated with
the yields of \cite{tny86} assuming a delay-time distribution function
in the form of a power law with an index of $-1$ (see
\citealt{mmn14}).  We refer to \cite{c15} for more information about
the treatment of stellar yields in our chemical evolution code.

We deliberately do not consider the mass ejected by stars more massive
than $30\,$M$_\odot$.  Some of the yields used in our work possess
islands of non-explodability, and we did not want to complement our
set of yields with other models that do not show this feature, as they
could bias and hide the importance those islands in our analysis.
But, such massive stars can eject a significant amount of light
elements during their pre-supernova evolution in the form of winds or
eruptions (e.g., \citealt{h07,cl13}) and can therefore contribute to
the chemical evolution, in spite of the slope of the stellar initial
mass function.  The predictions for the mass loss of the most massive
stars are however rather uncertain (e.g., \citealt{h03}) and possibly
significantly affected by binary star evolution.  For that reason, our
numerical predictions are probably underestimating the abundances of
certain elements such as O and Mg.  For the purpose of this paper,
however, this is not a limitation because we are only interested in
the differential changes due to assumptions in the mass and
metallicity resolution of the yield grid.

\section{Stellar Abundances Data}
\label{sect_sad}
To provide a visual reference of the impact of the grid resolution of stellar yields,
we compare our results with the stellar abundances observed in the Milky Way
for O, Na, Mg, Si, Ca, Ti, and Mn.  The data has been plotted using the STELLAB
module, which stands for STELLar ABundances.  This python code is also available online
with the NuGrid NuPyCEE package.  It uses a stellar abundances database and
plot any abundance ratio for the Milky Way, Sculptor, Carina, Fornax, and the
Large Magellanic Cloud.  It should be stressed that the database is
not curated and for now consist of a collection of data that has been blindly taken from the literature.
For this work, however, we only took a sample of the entire STELLAB Milky Way database
to provide a cleaner and more representative view of the global chemical evolution trends.
Although our data selection includes the Galactic halo and thick and thin discs, we remind
the reader that we use a one-zone model and do not consider those three components independently.

In our sample, there is no star duplication at [Fe/H] above $-2$, as all data in this metallicity
range come from only one source, which is either \cite{bfo14} for O, Na, Mg, Si, Ca, and Ti,
or \cite{bb15} for Mn.  At lower [Fe/H], the stellar abundances mostly come from \cite{c04} and
\cite{c13} and may contain star duplications, except for O for which the data only come
from \cite{c04}.  There is a lack of data around [Fe/H]~$\sim-2$ in our selection.  We
could have complemented our selection with other studies like \cite{i12,i13}, but we decided
to limit the amount of data to improve the clarify of our figures and to make the reading
of our numerical predictions easier.  This data gap should then be considered as
a selection bias.

At low [Fe/H], O abundances have been derived using the [\ion{O}{i}]\,$\lambda$6300 line
along with a correction for 3D effects, while at high [Fe/H], O abundances have been
derived using the \ion{O}{i}\,$\lambda$7774 triplets along with a correction for NLTE effects.  
We excluded carbon-enhanced metal-poor stars from the \cite{c13} dataset
to better isolate the global chemical evolution trends, which represent the best target for one-zone models.
Na abundances can significantly be affected by LTE departure, especially at low [Fe/H]
(e.g., \citealt{j15}).  Because \cite{c13} did not include NLTE effects for Na,  
we replaced those data by the NLTE-corrected Na abundances of \cite{r14}.
All data and numerical predictions presented in the following figures are normalized to the solar
abundances found in \cite{a09}.

\begin{figure}
\begin{center}
\includegraphics[width=3.54in]{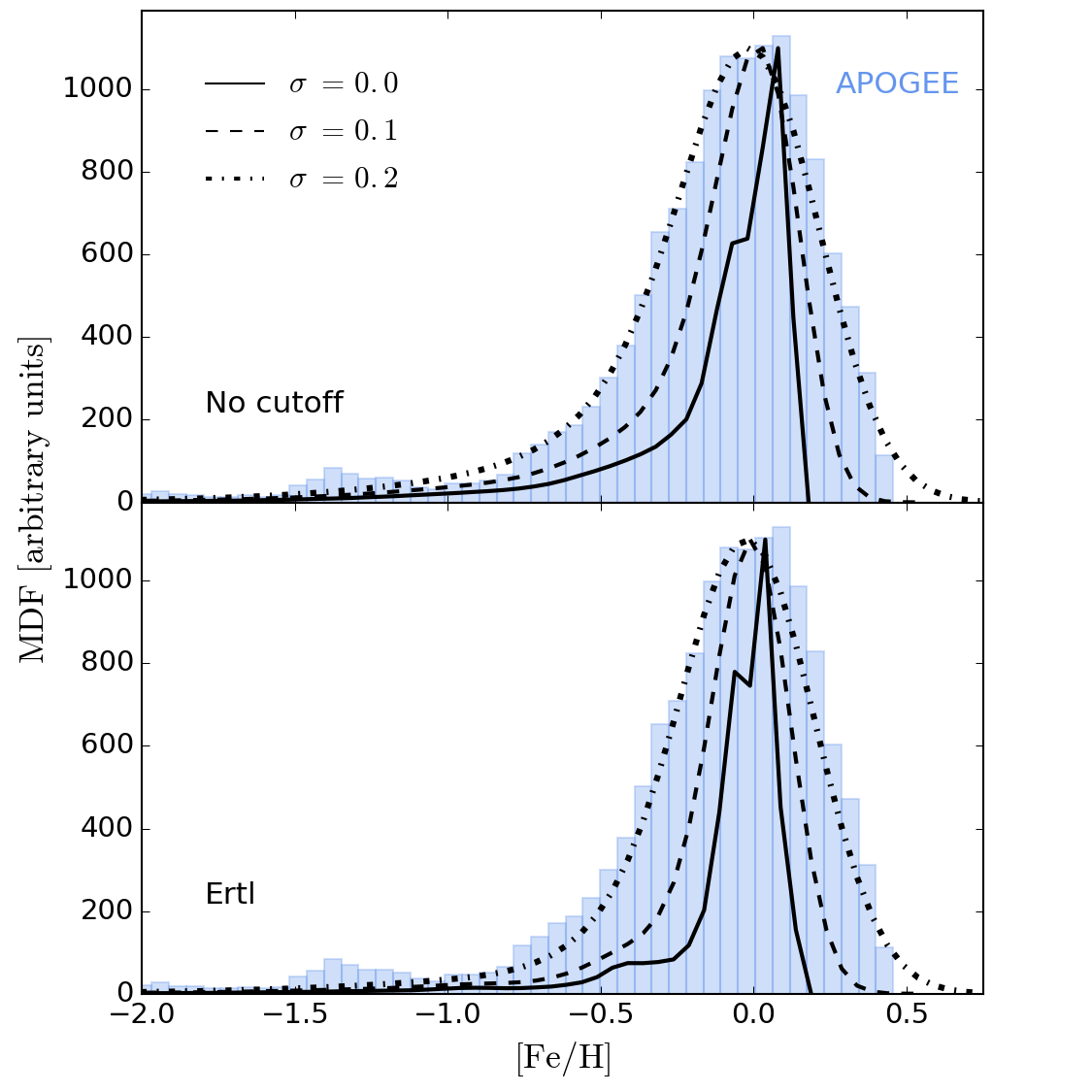}
\caption{Predicted metallicity distribution function generated using stellar yields with the no-cutoff (upper panel) and the \protect\cite{e15} (lower panel) remnant mass prescriptions.  The solid line represents the raw output extracted from our one-zone model, while the dashed and dot-dashed lines represent a convolution between the raw output and gaussian functions with a standard deviation of 0.1 and 0.2, respectively.  The blue histogram has been extracted from the APOGEE R12 dataset for the Milky Way.}
\label{MDF}$ $
\end{center}
\end{figure}

\begin{figure*}
\begin{center}
\includegraphics[width=6.9in]{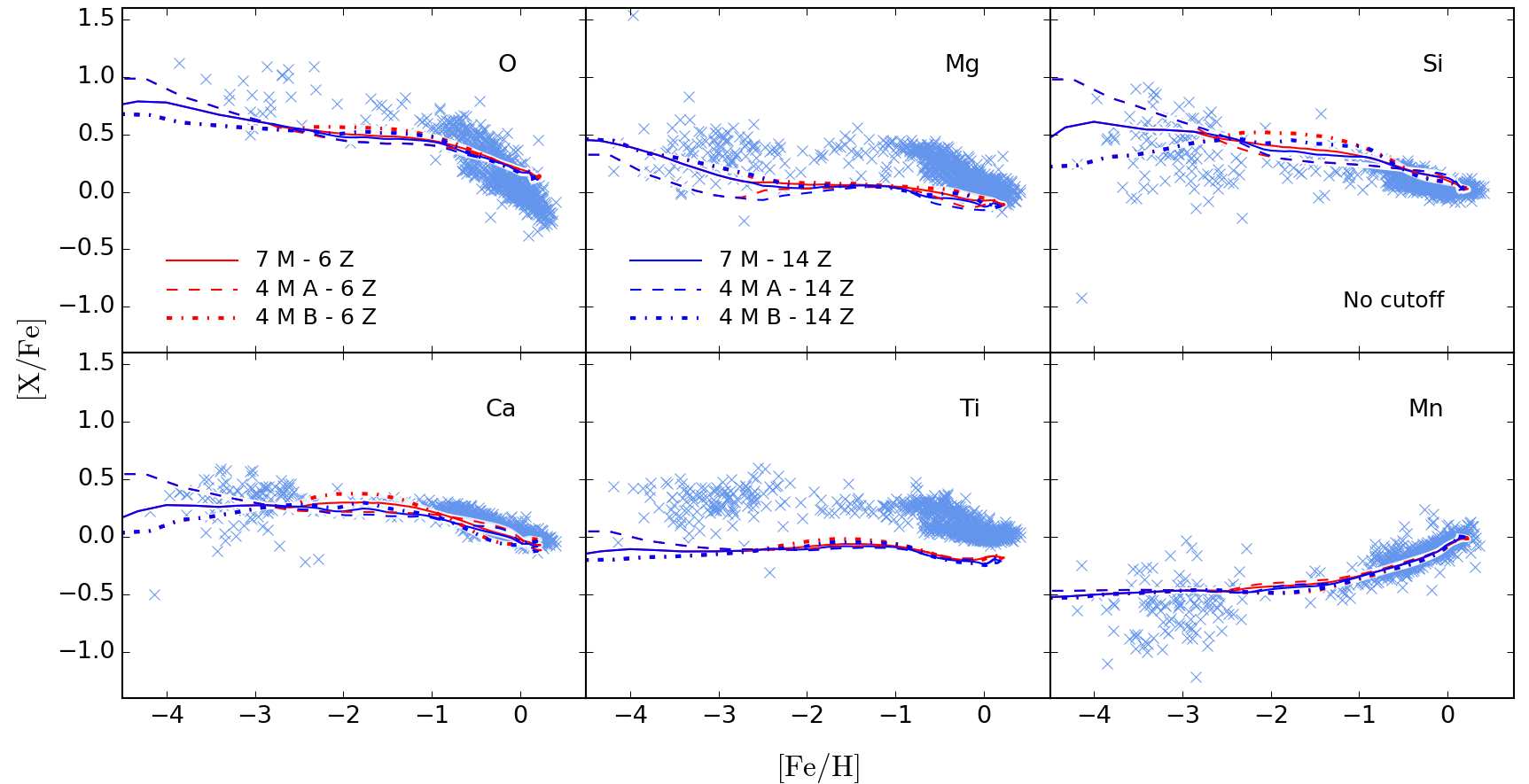}
\caption{Impact of the metallicity resolution for different mass
  resolutions on the predicted evolution of six elements, relative
  to Fe, as a function of [Fe/H] using the no-cutoff prescription for
  the remnant mass of massive stars.  The nomenclature associated with
  the different line styles and colours is described in
  Table~\ref{tab_sample}.  Observational data come from 
  \protect\cite{c04}, \protect\cite{c13}, \protect\cite{bfo14}, \protect\cite{r14},
  and \protect\cite{bb15} (see Section~\ref{sect_sad}).}
\label{X_Fe_no_Z_res}
\end{center}
\end{figure*}

\begin{figure}
\begin{center}
\includegraphics[width=3.54in]{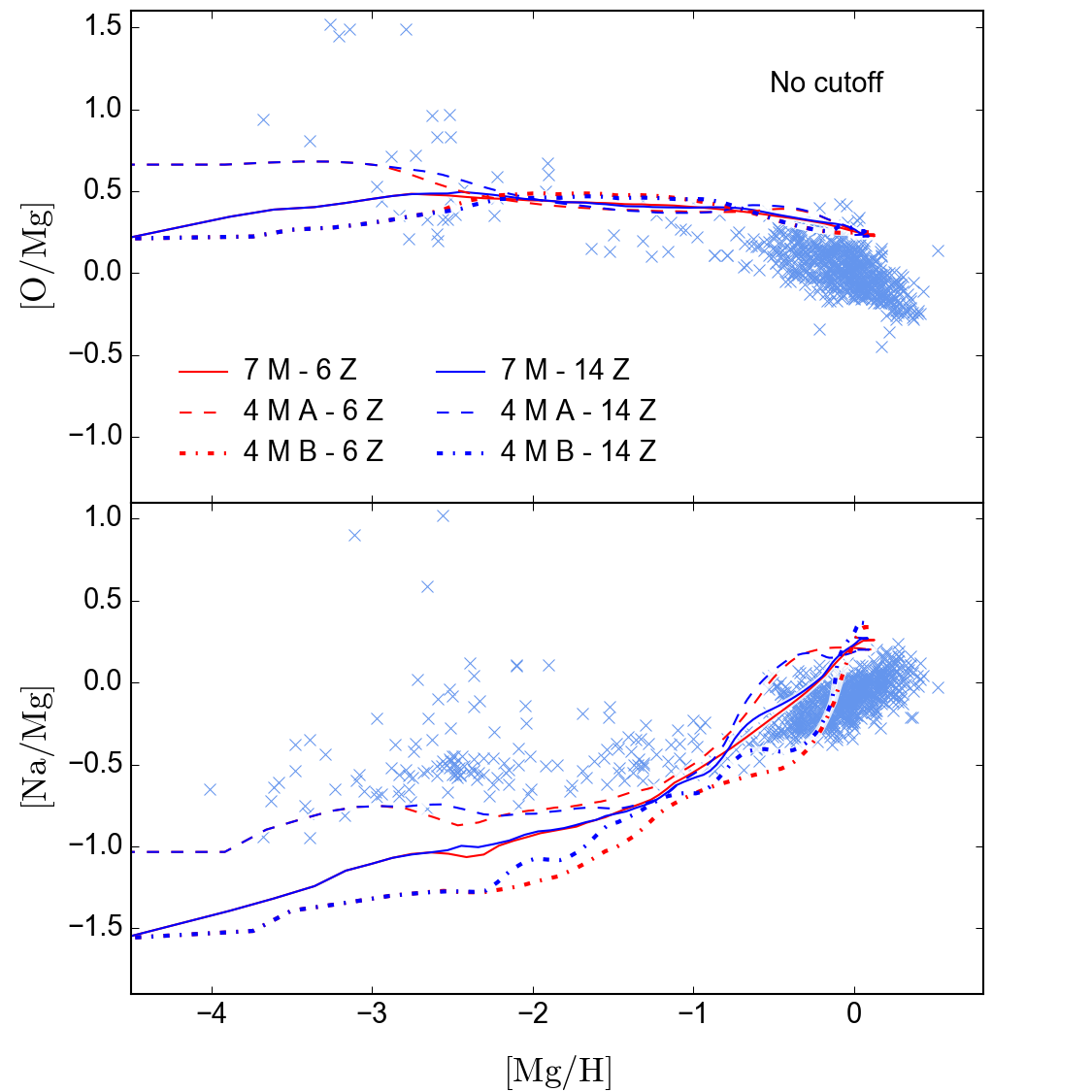}
\caption{Impact of the metallicity resolution for different mass
  resolutions on the predicted evolution of O and Na, relative to Mg,
  as a function of [Mg/H] using the no-cutoff prescription for the
  remnant mass of massive stars.  The lines and the observational data
  are the same as in Figure~\ref{X_Fe_no_Z_res}.}
\label{O_Na_Mg_no_Z_res}$ $
\end{center}
\end{figure}

\section{Yields with Monotonic Remnant Masses}
\label{sec_y_mrm}
In this section, we explore the impact of different mass and
metallicity resolutions on galactic chemical evolution predictions,
using the no-cutoff prescription for the remnant mass of massive
stars.  Throughout the following figures, we use the nomenclature
defined in Table~\ref{tab_sample} to label which stellar models were
sampled from the original grid.  As mentioned in Section~\ref{sect_yields}, the
stellar yields at $[Z]=0.3$ are not considered, since the final metallicity
in our numerical predictions typically does not exceed [Fe/H]$\sim0.2$.
With the \textit{5 Z} and \textit{6 Z} samples, we use the yields at $[Z]=0.1$
when the composition of the galactic gas exceeds this metallicity.
The adopted values for some of our key parameters as well as
the final properties of our Milky Way models
are given in Table~\ref{tab_final_prop}.

Our predicted metallicity distribution
function (MDF), for the present choice of yields, is shown in the upper
panel of Figure~\ref{MDF} and compared with the MDF we extracted
from the APOGEE R12 dataset (\citealt{h_ap15,s_ap15,g_ap16}).  We
chose APOGEE to maximize the statistics without having star duplication.
To broaden our MDF, we convolved it with gaussian functions
using different values for the standard deviation parameter, $\sigma$.
In our case, this serves to mimic non-uniform mixing and stochastic processes 
that are not included yet in our one-zone model.  
This convolution process has been used before (e.g., \citealt{pg12,cote13,pilk13}) and shows that, with additional scatter,
our predicted MDF could be in reasonable agreement with observations.
But our narrow raw MDF (solid line in Figure~\ref{MDF})
implies that our model currently does not capture the full complexity of 
the formation of the Milky Way, even if our model reproduces
its current global properties (see Table~\ref{tab_final_prop}).
In particular, our model is probably not suited to reproduce the
early evolution of the Galactic halo.  But given the purpose
of the present study, we still start our simulations with primordial gas in 
order to cover the metallicity range included in our
stellar yields.

\begin{figure*}
\begin{center}
\includegraphics[width=6.9in]{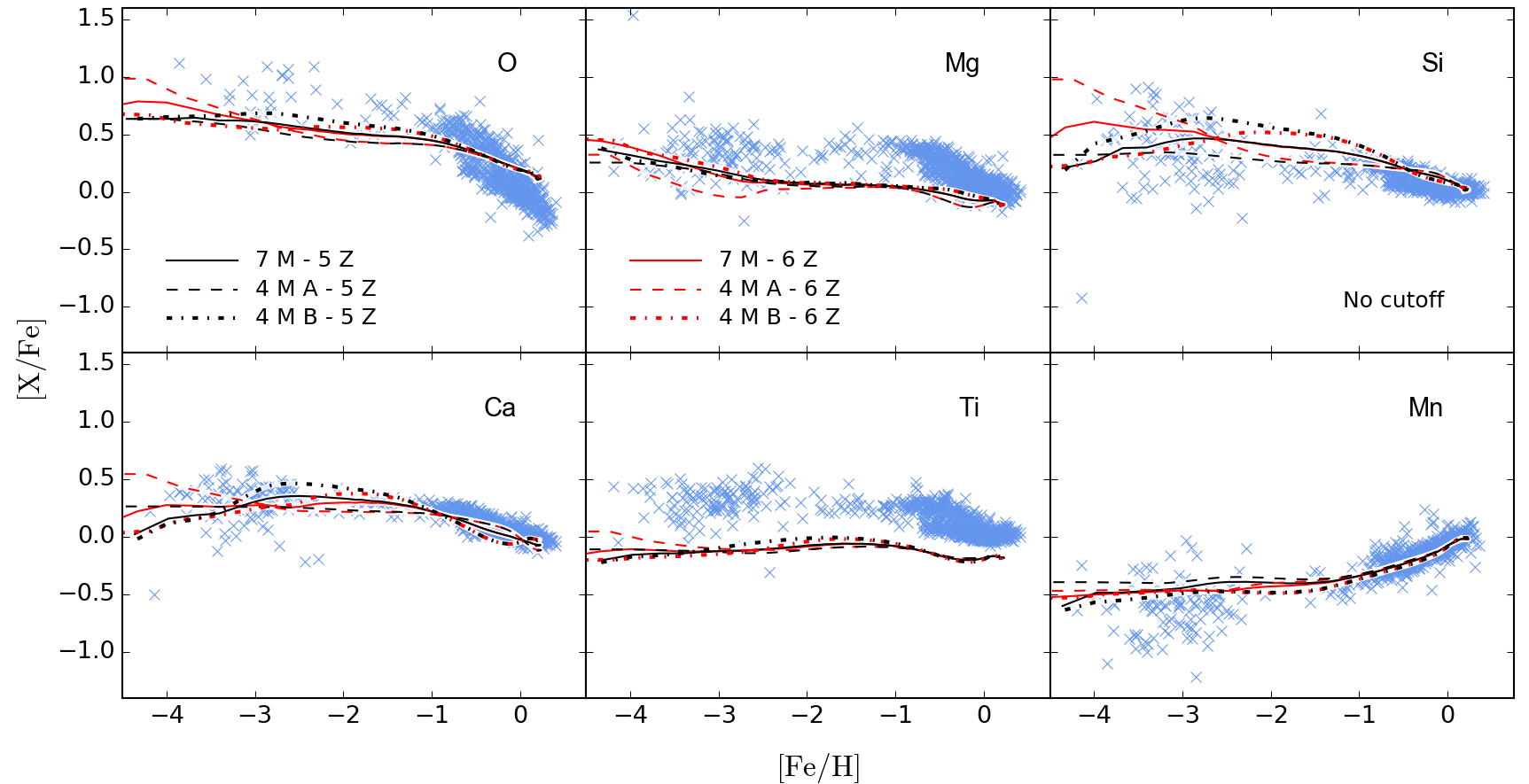}
\caption{Impact of the metallicity range for different mass
  resolutions on the predicted evolution of six elements, relative
  to Fe, as a function of [Fe/H] using the no-cutoff prescription for
  the remnant mass of massive stars.  The nomenclature associated with
  the different line styles and colours is described in
  Table~\ref{tab_sample}.  The observational data are the same as in
  Figure~\ref{X_Fe_no_Z_res}.}
\label{X_Fe_no}
\end{center}
\end{figure*}

\subsection{Mass and Metallicity Resolutions}
Figure~\ref{X_Fe_no_Z_res} presents our predictions for six
elements, relative to Fe, against the stellar abundances observed in
the Milky Way.  For a given colour, the different line styles
illustrate the impact of using different mass samplings and
resolutions.  For a given line style, the red and blue lines
illustrate the impact of using different metallicity resolutions.
With all the 14 metallicities sampled (blue lines), this last figure
shows that different mass samplings produce different results when the
number of masses is reduced to four (dashed and dot-dashed lines).  At
[Fe/H]~$\lesssim-2$, with the stellar yields used in this work,
different selections of masses generate variations of about $0.2-0.3$~dex
for O, Mg, and Ti, and $0.5-0.7$~dex for Si and Ca.  At higher [Fe/H], our predictions are less sensitive
to the selection of masses as variations are generally found within
0.1~dex.  Within this [Fe/H] range, Ti and Mn are relatively insensitive to the mass resolution.

The variations seen at low [Fe/H] also highlight the impact of the
mass range considered in the yields.  The Mass Sampling A and B
(see Table~\ref{tab_sample}) include a maximum stellar mass of 25 and
30\,M$_\odot$, respectively.  Those massive star models, for the lowest
metallicity included in the yields, are the first to enrich the galactic gas
at early time.  Below [Fe/H]~$\sim-3$, the predictions generated
by the Mass Sampling A and B (dashed and dot-dashed lines in Figure~\ref{X_Fe_no_Z_res})
therefore represent, respectively, the ejecta of the 25 and the 30\,M$_\odot$ models.
In fact, we made a test where we modified the Mass Sampling A to
include both the 25 and the 30\,M$_\odot$ models.  In that case,
for [Fe/H] $\lesssim-3$,  the dashed lines in Figure~\ref{X_Fe_no_Z_res} became similar
to the solid lines where all seven masses are sampled.
However, above [Fe/H] $\sim-3$, the variations between the different mass
resolutions and samplings remained unchanged.  This shows how
sensitive our numerical predictions at early time are to the selection
of the first stellar models that participate in the enrichment process.

\begin{figure}
\begin{center}
\includegraphics[width=3.54in]{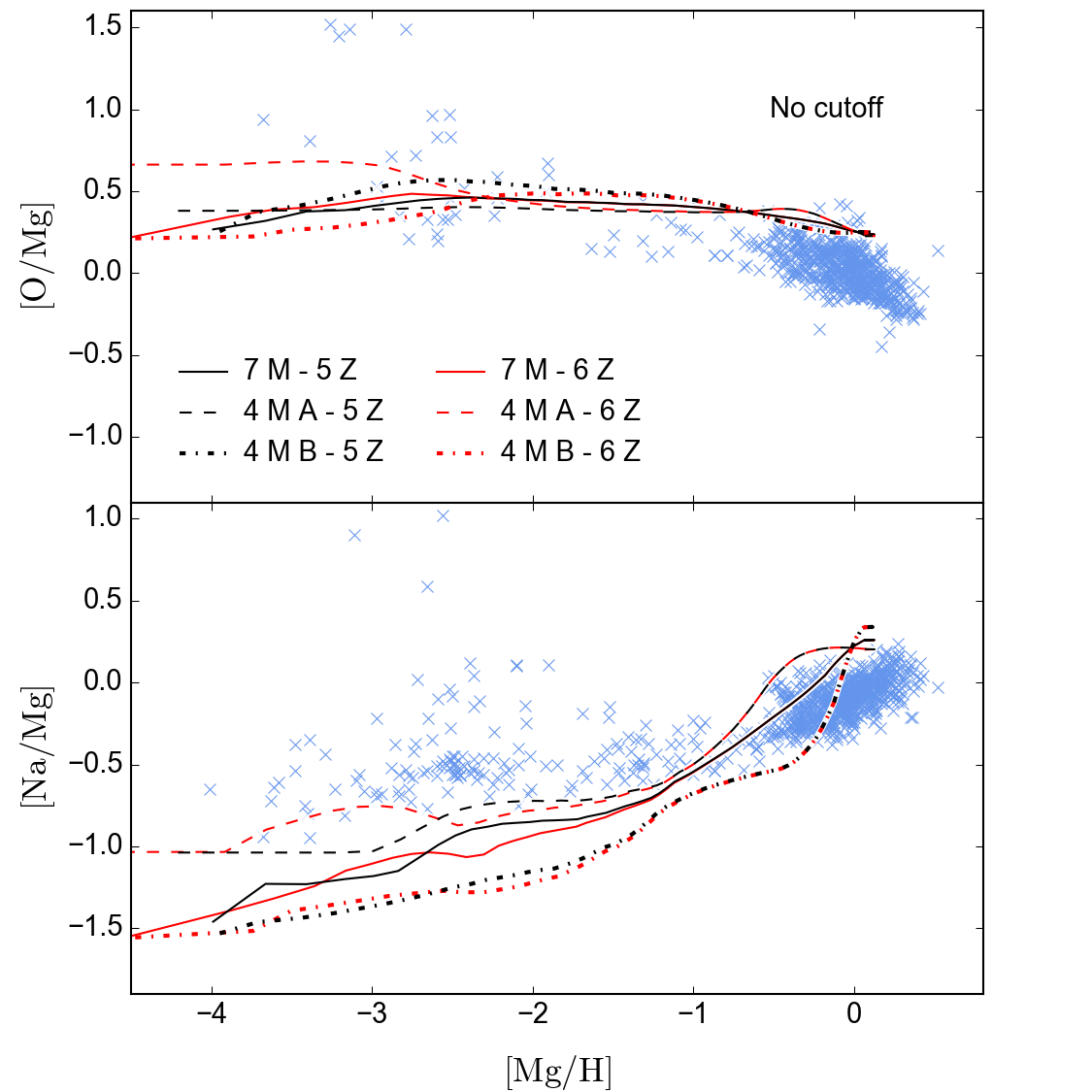}
\caption{Impact of the metallicity range for different mass resolutions
  on the predicted evolution of O and Na, relative to Mg,
  as a function of [Mg/H] using the no-cutoff prescription for the
  remnant mass of massive stars.  The lines and the observational data
  are the same as in Figure~\ref{X_Fe_no_Z_res}.}
\label{fig_O_Na_Mg_no}$  $
\end{center}
\end{figure}

Reducing the number of metallicities does not produce any significant
change in our numerical predictions, as shown by the red and blue
lines in Figure~\ref{X_Fe_no_Z_res}.  Figure~\ref{O_Na_Mg_no_Z_res}
presents an analogous of Figure~\ref{X_Fe_no_Z_res} but for the
evolution of O and Na relative to Mg, as these three elements are all
mainly produced during the pre-supernova phases.  These predictions also converge toward
the idea that the mass resolution and sampling are more important than the
metallicity resolution.  Indeed, in the case of Na, the different mass
samplings generate more than 0.5~dex of variations at almost every
[Mg/H] value, whereas reducing the number of metallicities only
produce variations of about 0.1~dex.

We note that the variations
seen for elements that do not match observations, such as
Na and Ti, may not be representative, since we know stellar yields
need to be examined in more details.  However, we
still decided to show these elements to highlight where improvements
are needed in our stellar models.

\begin{figure*}
\begin{center}
\includegraphics[width=6.9in]{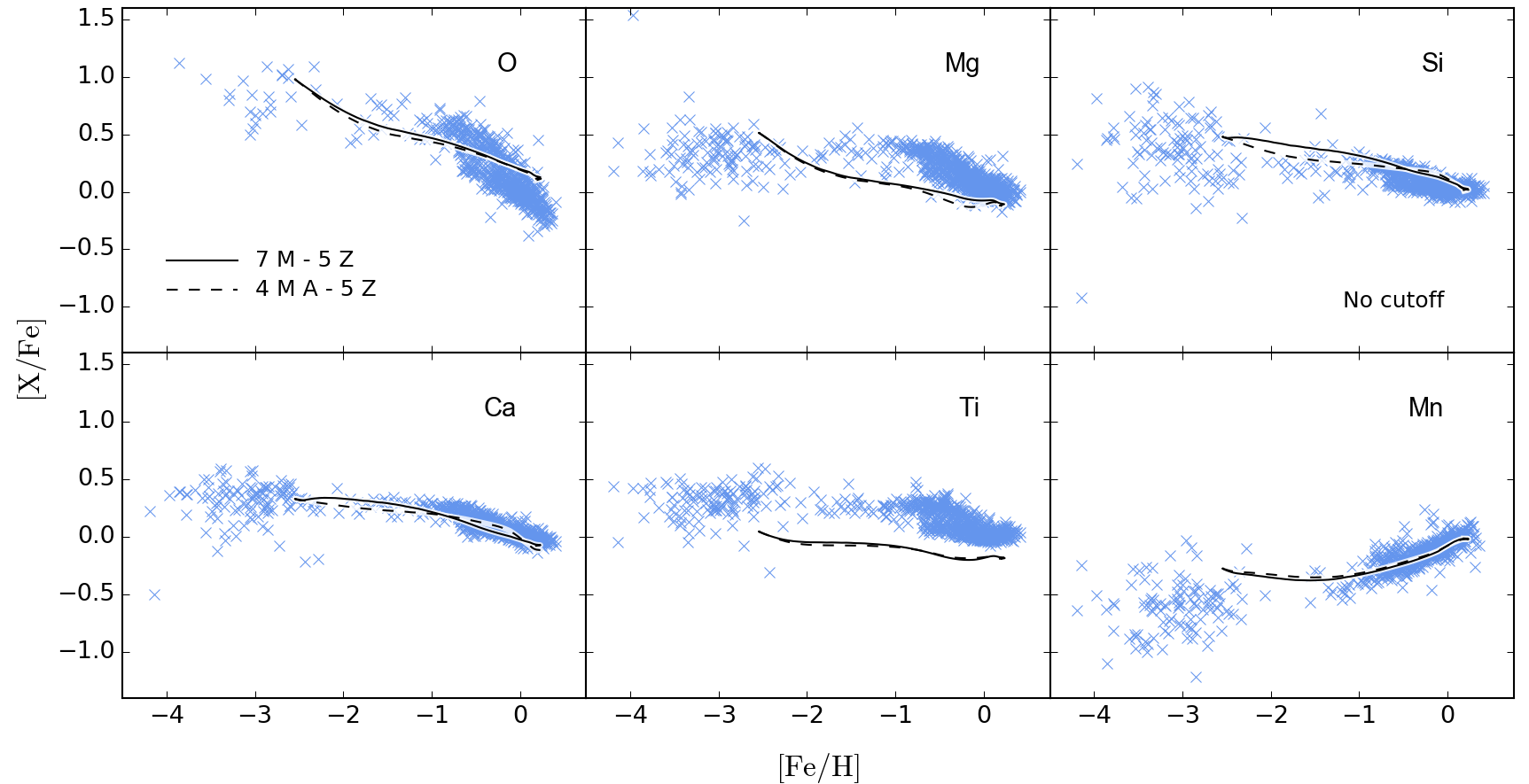}
\caption{Impact of using seven masses instead of four (Mass Sampling A) on the predicted evolution of six elements, relative to Fe, as a function of [Fe/H] using the no-cutoff prescription for the remnant mass of massive stars with five metallicities ranging from $[Z]=-2$ to 0.1.  The simulations have been calculated with an initial gas composition set to $Z=0.000153$.  The observational data are the same as in Figure~\ref{X_Fe_no_Z_res}.}
\label{X_Fe_NuGrid}
\end{center}
\end{figure*}

\subsection{Mass Resolution and Metallicity Range}
Figure~\ref{X_Fe_no} shows the impact of using $[Z]=-2$ instead of
$Z=0$ as the lower boundary of the metallicity range covered by our
set of yields.  Because of the wide range of considered metallicities, 
our yields interpolation scheme between the sampled metallicities 
is done in the log space.  This, unfortunately, prevents us from including $Z=0$ 
in the interpolation, as its logarithm value is not a finite number. Therefore,
when the zero-metallicity yields are included, we only
use them for stars formed in primordial gas, and switch to
$[Z]=-2$ as soon as the gas gets enriched by the first stellar
ejecta.  In the case without the zero-metallicity yields, we use the
$[Z]=-2$ yields all the way until the metallicity of the gas
actually reaches $[Z]=-2$, above which we interpolate between
metallicities.\\
\indent When excluding the zero-metallicity yields (black
lines), the choice of the mass sampling has generally the biggest impact at
[Fe/H]~$\sim-2$, as opposed to [Fe/H]~$\lesssim-3.5$ in the case where the
zero-metallicity yields are included (red lines).  This suggests that
the lowest metallicity available in the yields plays a dominant role
in the numerical predictions at low [Fe/H].  As a matter of fact,
depending on the mass sampling, including or not the zero-metallicity
yields generally produces different results that do not converge
before reaching [Fe/H] of $\sim-2$.  The situation also 
occurs when looking at the evolution of O and Na as a function of [Mg/H]
(Figure~\ref{fig_O_Na_Mg_no}).  The lowest [Fe/H] and [Mg/H] values
of our numerical predictions correspond to the first timestep that includes
chemical enrichment, and depend on the amount of Fe and Mg ejected
by the most massive stellar model sampled in the set of yields, for the
lowest metallicity.  When excluding the zero-metallicity yields (black lines), the
relatively small variations at [Fe/H]~$\lesssim-3$, especially for O, Mg, and Si,
are due to a similarity in the ejecta composition of the 25 and 30\,M$_\odot$ models
at $[Z]=-2$, which is the opposite in the models at $Z=0$.

\begin{figure*}
\begin{center}
\includegraphics[width=6.9in]{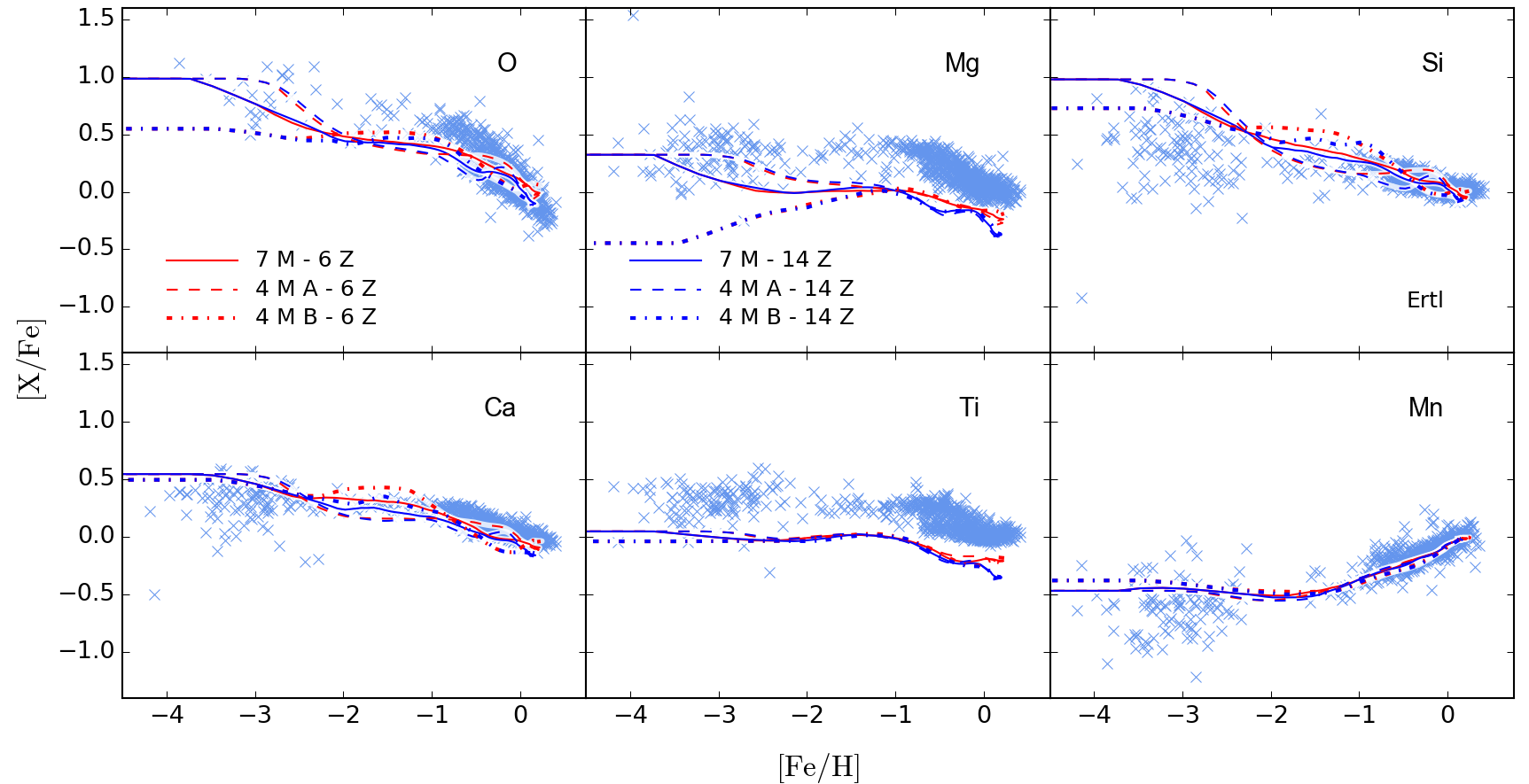}
\caption{Same as in Figure~\ref{X_Fe_no_Z_res} but with the remnant
  mass prescription of \protect\cite{e15}.}
\label{X_Fe_ertl_Z_res}
\end{center}
\end{figure*}

\begin{figure*}
\begin{center}
\includegraphics[width=6.9in]{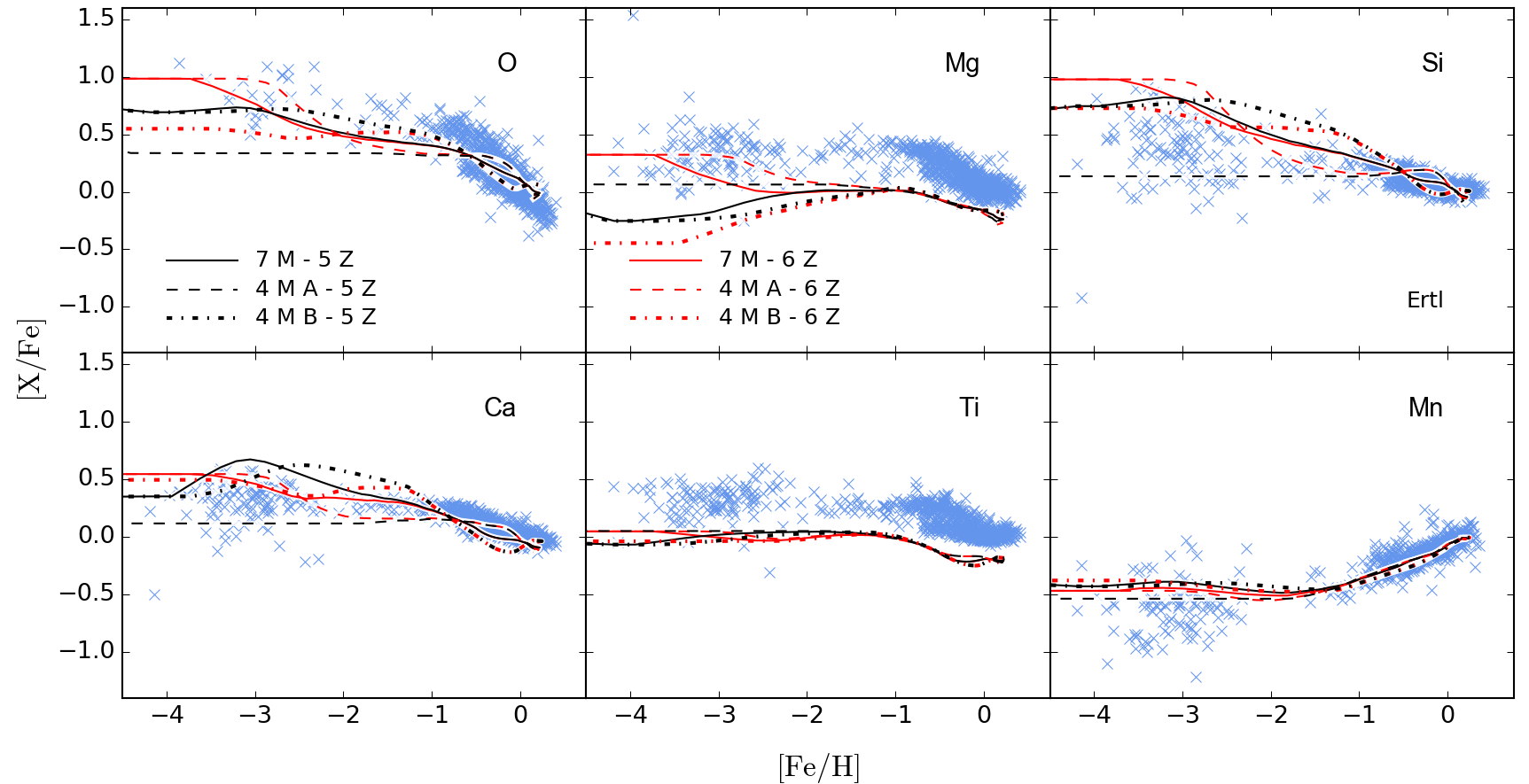}
\caption{Same as in Figure~\ref{X_Fe_no} but with the remnant mass
  prescription of \protect\cite{e15}.}
\label{X_Fe_ertl}
\end{center}
\end{figure*}

\subsection{Number of Masses in NuGrid Yields}
NuGrid stellar yields currently include five metallicities ($Z=0.02$, 0.01, 0.006, 0.001, and 0.0001) and four models per metallicity for massive stars ($M=12$, 15, 20, and $25\,$M$_\odot$).  The question that ignited the present study was whether NuGrid should add more masses in their set of yields.  To answer that question, we assumed that the current state of NuGrid could be represented by the Mass Sampling A - 5~Z sample (see Table~\ref{tab_sample}) with the no-cutoff remnant mass prescription, as NuGrid does not for the moment consider islands of non-explodability.  Figure~\ref{X_Fe_NuGrid} illustrates what would happen if seven masses were used instead of four.

For this comparison, we did not start our simulations with primordial composition.  As shown in the previous sections, numerical predictions are significantly affected by the choice of stellar yields associated with the first stellar ejecta.  To eliminate this complication, and to only focus on the current metallicity range covered by NuGrid, we started our simulations with the gas composition calculated in \cite{wh13} for $Z=0.000153$.  Given this configuration, we conclude from Figure~\ref{X_Fe_NuGrid} that adding more masses is not a major concern for NuGrid.  But still, the ideal case would be to provide a finer grid that do not produce variations when a few models are removed from the set of yields.

\section{Yields with Islands of Non-Explodability}
\label{sect_y_ine}
In this section we repeat the experiment made in the previous section,
but using the set of yields generated with the remnant mass
prescription of \cite{e15}, which included islands of
non-explodability.  The adopted input parameters and our predicted MDF, for this choice of yields, are presented
in Table~\ref{tab_final_prop} and in the lower panel of Figure~\ref{MDF}, respectively.  The MDFs
generated with our two remnant mass prescriptions are roughly similar.
The minor differences are due to different Fe ejection rates which 
are affected by the number of exploding models at a given time.

\subsection{Mass and Metallicity Resolutions}
Figure~\ref{X_Fe_ertl_Z_res} shows the impact of the mass and
metallicity resolutions using the remnant mass prescription of
\cite{e15}.  At [Fe/H] below~$\sim-2.0$, the results with six and 14
metallicities are still indistinguishable.  At [Fe/H]~$\sim-1.5$, for
Mass Sampling B, the metallicity resolution only generates
variations of 0.1~dex at most.  At [Fe/H]~$\sim-0.5$, for Mass Sampling A, variations
are generally between 0.05~dex and 0.15~dex, whereas almost no variation
was seen with the no-cutoff prescription (see
Figure~\ref{X_Fe_no_Z_res}).

In the case of Si, Ca, and Ti, the impact of the mass resolution at
[Fe/H]~$\lesssim-3$ is less important than with the no-cutoff
prescription, as opposed to O and Mg which now show variations ranging
from 0.4~dex to 0.7~dex.  At [Fe/H] between $-2$ and $-1$, the impact of
the mass sampling is increased by about 0.1~dex relative to
Figure~\ref{X_Fe_no_Z_res}.  These results suggest that the mass
requirement in a set of stellar yields depends on the remnant mass
prescription and reinforce our conclusion that the mass resolution is
more important than the metallicity resolution.

\subsection{Mass Resolution and  Metallicity Range}
\begin{figure}
\begin{center}
\includegraphics[width=3.54in]{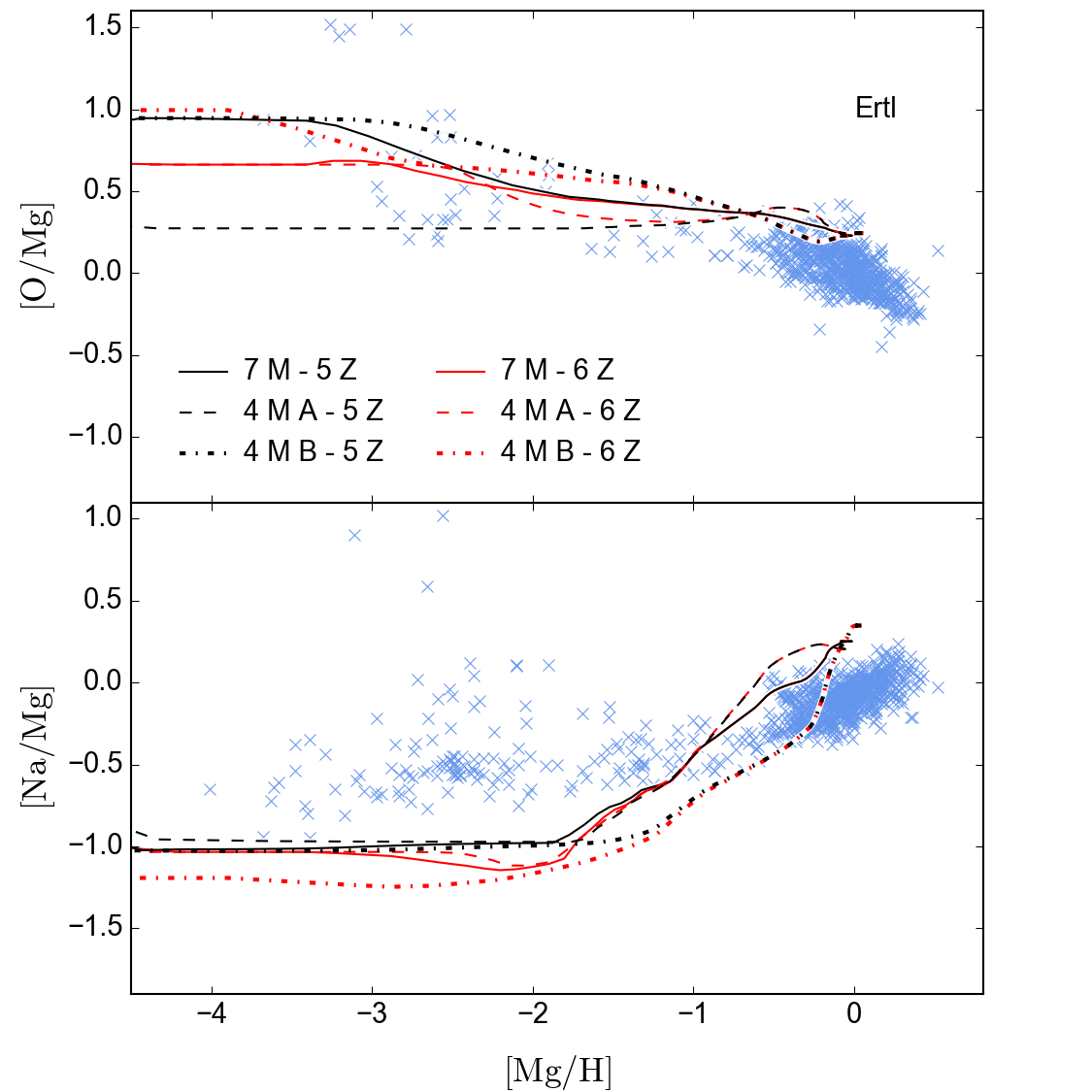}
\caption{Same as in Figure~\ref{fig_O_Na_Mg_no} but with the remnant
  mass prescription of \protect\cite{e15}.}
\label{fig_O_Na_Mg_ertl}
\end{center}
\end{figure}

\begin{figure}
\begin{center}
\includegraphics[width=3.54in]{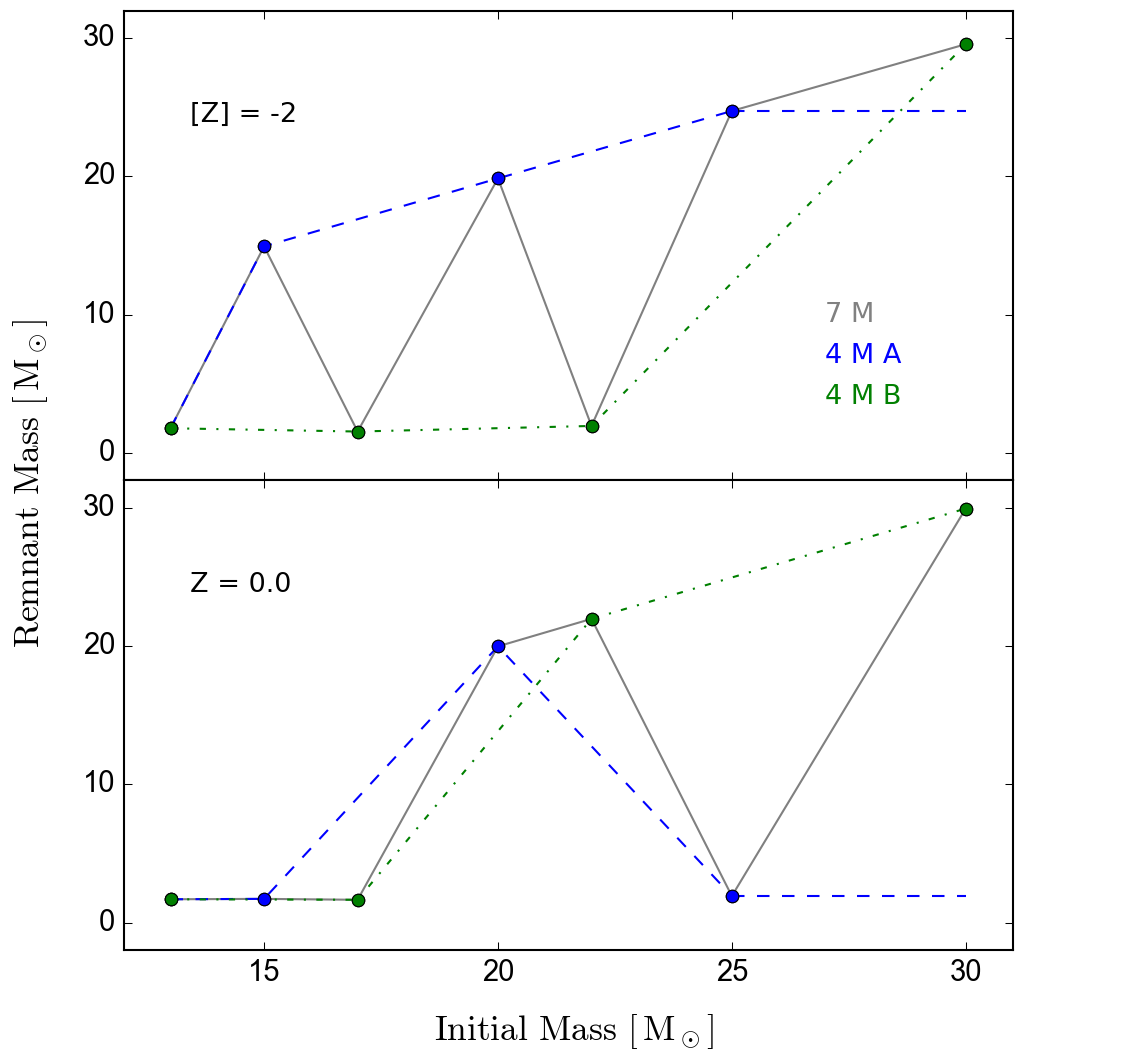}
\caption{Remnant mass as a function of initial stellar mass with the
  prescription of \protect\cite{e15} for two different metallicities.
  The different coloured lines represent the different mass samplings
  defined in Table~\ref{tab_sample}.}
\label{remnant_2Z}
\end{center}
\end{figure}

The first feature to notice in Figure~\ref{X_Fe_ertl} is the
importance of the mass sampling when only five masses and five
metallicities between 0.000153 and 0.0193 are considered (black dashed
and black dash-dotted lines).  With the stellar yields used in this
work, this can generate variations up to $0.3-0.4$ dex for O and Mg,
and up to $0.5-0.7$ dex for Si and Ca across a larger [Fe/H]
interval than in the case with the no-cutoff prescription.  Ti and Mn are
still insensitive to the mass sampling and resolution.  Above [Fe/H]
$\sim$ $-0.5$, all predictions are relatively
insensitive to the mass resolution compared to the variations seen at low [Fe/H].  When plotted relative to Mg (see
Figure~\ref{fig_O_Na_Mg_ertl}), O shows variations up to 0.6~dex at
[Mg/H] below $\sim$~$-2.0$, whereas Na shows variations up to 0.4~dex
at [Mg/H] above $\sim$~$-1.0$.\\
\indent These significant variations are explained by the upper panel of
Figure~\ref{remnant_2Z}.  At $[Z]=-2$, the islands of
non-explodability are regularly dispersed across the initial stellar
mass.  With the Mass Sampling A, the whole selection consists of
non-exploding models, except for the $13\,$M$_\odot$ model.  On the other
hand, the Mass Sampling B selects all the explosive models and misses
the islands of non-explodability located at $15\,$M$_\odot$,
$20\,$M$_\odot$, and $25\,$M$_\odot$.  Therefore, when the stellar models at
$[Z]=-2$ are the first to enrich the primordial gas, the mass
resolution is crucial.  The situation is less extreme at $Z=0$ (see
the lower panel of Figure~\ref{remnant_2Z}).  But still, in order to
resolve the islands of non-explodability, the stellar models must be
judiciously sampled.  For example, one could have a representative
selection without the $15\,$M$_\odot$ and $20\,$M$_\odot$ models, but not
without the $25\,$M$_\odot$ model.

The different magnitudes of the impact of the mass sampling with and
without the zero-metallicity yields, especially for Si and Ca,
reinforces the idea that the lowest metallicity included in our set of
stellar yields plays a major role in the numerical predictions at low
[Fe/H].  In this case, even when all the seven masses are considered
(solid lines), modifying the lowest metallicity available produces
different results that do not converge before reaching
[Fe/H]~$\sim$~$-3$ for O and Si, [Fe/H]~$\sim-2.5$ for Mg,
and [Fe/H]~$\sim$~$-1.5$ for Ca.  This is consistent with observations and simulations
that show a rapid increase of the average metallicity at early time in
the Milky Way (e.g., \citealt{kn11,bfo14}).

\begin{figure*}
\begin{center}
\includegraphics[width=6.9in]{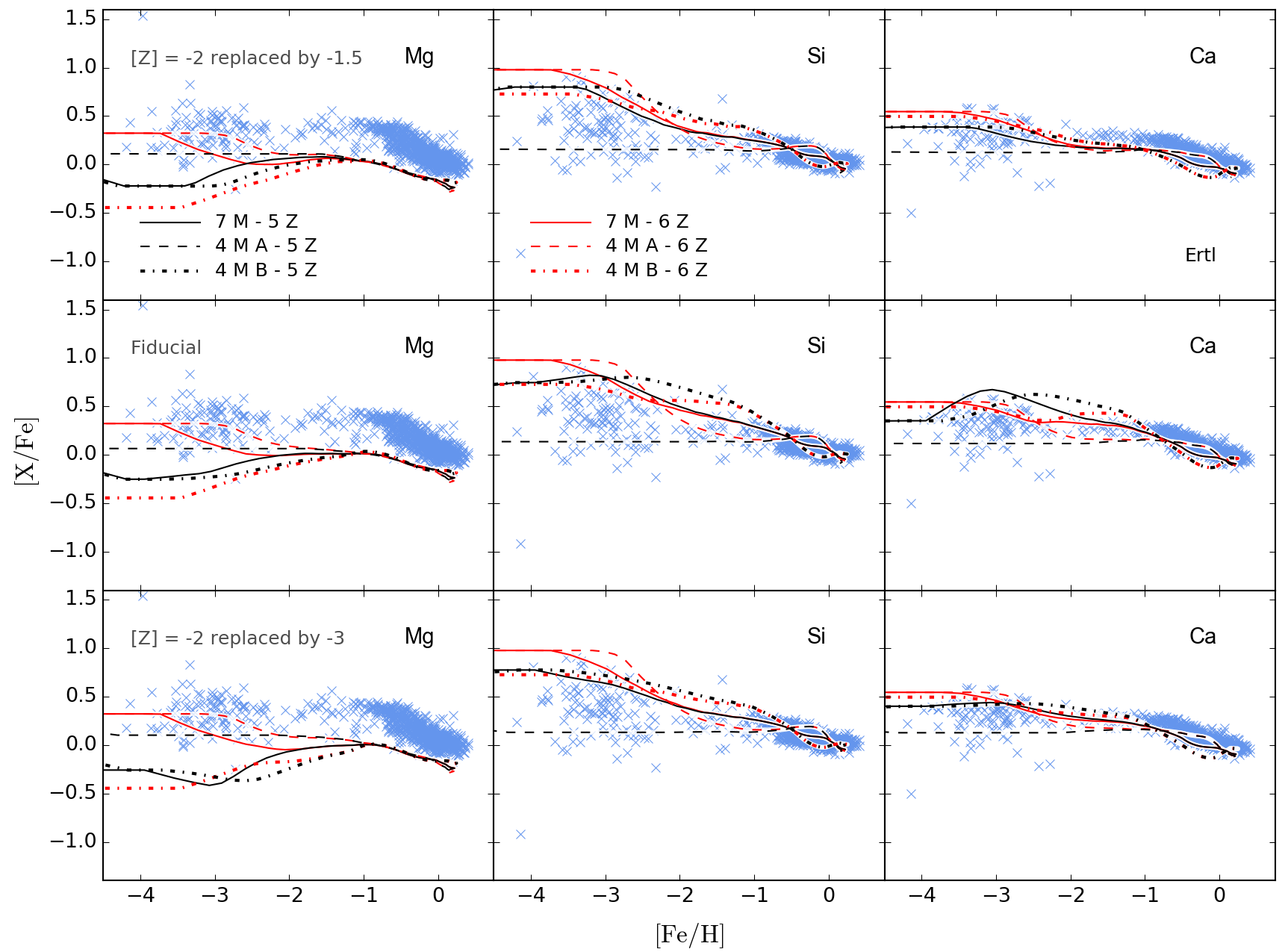}
\caption{Same as in Figure~\ref{X_Fe_ertl} for Mg, Si, and Ca, but with different choices for the lowest non-zero metallicity in the sample.}
\label{fig_1st_nonzero_Z}
\end{center}
\end{figure*}

Figure~\ref{fig_1st_nonzero_Z} shows an analogous of Figure~\ref{X_Fe_ertl} 
where we replaced $[Z]=-2$, the lowest non-zero metallicity in our fiducial
case (middle panels), by $[Z]=-3$ (lower panels) and $[Z]=-1.5$ (upper panels).
The predictions generated using the zero-metallicity
yields (red lines) are still similar from one case to another below [Fe/H]~$\sim-2.5$, while variations
between cases (upper, middle, and lower panels) can be seen between [Fe/H]~$\sim-2$ and $-1$.  When the 
zero-metallicity yields are not included (black lines), variations can be seen up to [Fe/H]~$\sim-1$
between the different cases, especially for Ca.  The
black dashed lines are always flat and similar from one case to another since the Mass
Sampling A only selects one explosive model at $[Z]=-3$, $-2$, and $-1.5$ (see Figure~\ref{fig_2D_remnant}).
Results shown in Figure~\ref{fig_1st_nonzero_Z} indicate once more that, below [Fe/H]~$\sim-2$,
our numerical predictions are sensitive to the first stellar models that enrich the galactic gas.
However, we recall that we use a one-zone model that is
not necessarily suited to reproduce the Galactic halo, which
is associated with the metallicity range where most of the
variations occur.\\
\indent It is worth remembering
that the two sets of yields used in the present work have been
calculated in the same way.  The only ingredient responsible for the
differences between Figures~\ref{X_Fe_no} and \ref{X_Fe_ertl} is the
remnant mass prescription, which suggests again that the mass
resolution required in stellar yields depends on the remnant mass
prescription used for massive stars.  The stellar yields used in our experiment are not general.  The fact
that the extreme case of $[Z]=-2$ is masked by the
zero-metallicity yields (see the comparison between the red and black lines 
in Figure~\ref{X_Fe_ertl}) does not mean the solution is to always use
zero-metallicity yields.  Depending on the modeling assumptions, the
stellar evolutionary code, and the physics included, other sets of
yields could have different islands of explodability for their
zero-metallicity models (e.g., \citealt{s15,e15}).  In that case, the
choice of the mass sampling could have different repercussions than
the ones shown in this section for the cases including zero-metallicity
yields.\\
\indent The variations seen in our figures illustrate the potential of
our stellar yields to reproduce some of the observed scatter
at low [Fe/H], especially for O, Na, Mg, Si, and Ca.  In inhomogeneous-mixing
chemical evolution models (e.g., \citealt{g03,arg04,ces15,w15}), where the stellar initial mass function
is randomly sampled, individual stellar models can transfer their 
ejecta into new generations of stars without being covered up
by all other stellar models that would be contributing when the 
initial mass function is assumed to be fully sampled.
The level of scatter should
depends on the variety of abundance ratios seen in the ejecta of
the adopted stellar models.  We are currently working on a stochastic version of our chemical evolution
codes to generate scatter in our predictions.  This, however, is beyond
the scope of the present paper.

\section{Summary and Conclusion}
\label{sect_s_c}
We used a single-zone model to address the question of how many masses
and metallicities are needed in a grid of stellar yields in order to
generate relevant and reliable predictions with chemical evolution
models.  Using the set of stellar yields described in
Section~\ref{sect_yields}, which has seven masses and 15 metallicities
for massive stars, we performed experiments where we extracted a
subset of models to evaluate the impact of the grid resolution on the
chemical evolution of seven elements.  As a visual reference, we compared
our results with the stellar abundances observed in the Milky Way to
better appreciate the variations between our results.  Our work
suggests that there is no general answer to how many masses and
metallicities are need for galactic chemical evolution applications.\\
\indent The mass resolution needed in stellar yields depends on the element
considered and on the remnant mass prescription used for massive
stars.  We found that yields with a monotonic remnant mass
distribution are generally more robust to modifications in the grid
resolution.  Yields that possess islands of non-explodability are more
vulnerable, however, as explosive or non-explosive mass regimes can be
missed if not enough models are sampled.  Our results suggest that the
yields from the lowest metallicity included in the grid can dominate
the chemical evolution up to [Fe/H]~$\sim-2$.  Depending on the
remnant mass distribution applied for the lowest metallicities, a bad
mass sampling in the presence of islands of non-explodability can
cause variations that exceed 0.5~dex (see Figures~\ref{X_Fe_no} and \ref{X_Fe_ertl}).
The set of yields used in this work is not a general case and islands
of non-explodability could be found at different mass regimes in other
yields.  Under different stellar modeling assumptions, it is not
excluded that extreme cases, such as the one at $[Z]=-2$ with the \cite{e15}
prescription (see Figure~\ref{remnant_2Z}), could be associated with
other lower-metallicity or zero-metallicity yields.

We also studied the impact of the metallicity resolution and found
that a wide range is more important than the number of metallicities.
As for the mass resolution, yields with monotonic remnant masses are
less affected by the metallicity resolution, reinforcing our
conclusion that the grid resolution required in stellar yields depends
on the remnant mass prescription and on the presence or absense of
islands of non-explodability.  Similar results likely would be found
for any major discontinuities in yields as a function of initial mass.

\section*{acknowledgments}
We are thankful to Anna Frebel for relevant discussions on stellar abundance observation.
This research is supported by the National Science Foundation (USA) under
Grant No. PHY-1430152 (JINA Center for the Evolution of the Elements),
and by the FRQNT (Quebec, Canada) postdoctoral fellowship program.
AH was supported by an ARC Future Fellowship (FT120100363).  BWO was supported
by the National Aeronautics and Space Administration (USA) through grant
NNX12AC98G and Hubble Theory Grant HST-AR-13261.01-A.  He was also
supported in part by the sabbatical visitor program at the Michigan
Institute for Research in Astrophysics (MIRA) at the University of
Michigan in Ann Arbor, and gratefully acknowledges their hospitality.
FH acknowledges support through a NSERC Discovery Grant (Canada).
SB acknowledges support by JINA (ND Fund \#202476).

\label{lastpage}

\end{document}